# Optimization of the Block-level Bit Allocation in Perceptual Video Coding based on MINMAX

Chao Wang, Xuanqin Mou*, *Member, IEEE,* Lei Zhang, *Member, IEEE*

*Abstract*—In video coding, it is expected that the encoder could adaptively select the encoding parameters (e.g., quantization parameter) to optimize the bit allocation to different sources under the given constraint. However, in hybrid video coding, the dependency between sources brings high complexity for the bit allocation optimization, especially in the block-level, and existing optimization methods mostly focus on frame-level bit allocation. In this paper, we propose a macroblock (MB) level bit allocation method based on the minimum maximum (MINMAX) criterion, which has acceptable encoding complexity for offline applications. An iterative-based algorithm, namely maximum distortion descend (MDD), is developed to reduce quality fluctuation among MBs within a frame, where the Structure SIMilarity (SSIM) index is used to measure the perceptual distortion of MBs. Our extensive experimental results on benchmark video sequences show that the proposed method can greatly enhance the encoding performance in terms of both bits saving and perceptual quality improvement.

*Index Terms*—perceptual video coding, bit allocation, QP selection, block-level, MINMAX

## I. INTRODUCTION

IN lossy source encoding, the more bits an encoder uses to encode a source, the less distortion it derives, this is known as the rate-distortion (R-D) property [1]. In video encoding, a source can be a frame or a block. The R-D property of the sources in a video are usually quite different. Since the goal of video encoding is to convey video sequences at the best possible quality with limited bitrate, the encoder needs to choose the best scheme about how many bits should be used to encode each source, which is called optimization of bit allocation. The allocating of bits is controlled by the encoding parameters of the sources. In modern hybrid video encoding (e.g. H.264/AVC [2] and HEVC [3]) there are many parameters, such as quantization parameter (QP), block mode, and motion vector (MV), etc. Among these parameters, QP plays an important role in balancing the coding bitrate and quality, and most of the existing bit allocation algorithms focus on optimizing the QP selection [4-7]. The other parameters for a source is usually optimized after its QP is determined [8, 9], such as the R-D optimization of mode decision [10-13].

Optimizing the bit allocation (or QP selection) can be done at different coding levels (e.g. frame-level or block-level). Generally, the finer level the optimization processes on, the better performance it achieves. An optimized block-level bit allocation can not only improve the encoding performance on each frame, but also bring benefits to the whole sequence encoding, because a better reference frame can further improve the coding performance of the following frames [16]. However, there is hardly any method for optimal block-level bit allocation in practical encoding, most of the existing methods only do the optimization on frame-level. The reason for this is due to the high computational complexity in block-level bit allocation optimization, which is introduced by the dependency between blocks when intra and inter prediction is used.

When the sources are encoded independently, Lagrangian multiplier method [37] is usually used to solve the optimal bit allocation problem to achieve the minimum average distortion (MINAVE) [17] under given bit constraint. However, when the sources are dependent, the Lagrangian method could not be used directly because the R-D property of a source is unknowable before its reference sources are encoded. In this case, the dynamic programming (DP) method is usually used to solve the optimization problem [16-19], and its computational complexity has been proved to be increased exponentially with the number of dependent sources [17]. Unfortunately, the number of dependent blocks in a frame is usually so large that the complexity of the DP method for block-level optimization is unacceptable. Different from DP, some practical methods use models to predict the dependency of R-D property between adjacent frames, so that approximate optimal solutions of the frame-level bit allocation can be deduced before actual encoding [20-27], which can greatly reduce the encoding complexity. Nevertheless, these model-based methods are not suitable to be used in block-level optimization.

Some other works focus on minimizing the maximum distortion of sources (MINMAX). Since the human vision pays more attention to signals which have distinct difference with others [28], objects in a video or frame have much lower quality than others can court more attention. In this sense, the MINMAX criterion can bring benefits for perception. Many works have used MINMAX to reduce the quality fluctuation among frames [29-32]. The quality fluctuation inside one frame also should be reduced, [17] has shown the performance of MINMAX by an example of shape coding. Besides, based on MINMAX, one can use iterative-based method to approach the optimal solution with much lower complexity than DP [31].

Chao Wang is with the Institute of Image Processing and Pattern Recognition, Xi'an Jiaotong University, Xi'an, China (e-mail: cwang.2007@stu.xjtu.edu.cn).

Xuanqin Mou is with the Institute of Image Processing and Pattern Recognition, Xi'an Jiaotong University, Xi'an, China (e-mail: xqmou@mail.xjtu.edu.cn).

Lei Zhang is with the Dept. of Computing, The Hong Kong Polytechnic University, Hong Kong, China (e-mail: cslzhang@comp.polyu.edu.hk).

Unfortunately, there has hardly no practical MINMAX-based methods which can be used directly in block-level bit allocation. However, the virtues of MINMAX prompt us to construct a MINMAX-based block-level bit allocation method.

The metrics used to measure the distortion of blocks is critical in block-level bit allocation. Since the video content is shown to the human eyes, the perceived difference on distortion should be measured by perceptual-based metrics. Benefit by the great improvement of the research on the image quality assessment (IQA) technics, more and more IQA metrics, especially the Structure SIMilarity (SSIM) index [33], have been used in video encoding [10-15]. However, most of these works focus on improving the performance of mode decision, fewer works for optimizing the QP selection. Our previous works have used SSIM in a constant-quality-based method for block-level bit allocation [34, 35], and achieves promising results. However, more detailed works are needed to do in this aspect.

In this paper, we propose an MINMAX-based scheme to optimize the bit allocation (QP selection) at macroblock-level (MB-level) in H.264, SSIM is used to measure the distortion of MBs. An iteration-based maximum distortion descent (MDD) method is proposed to search the solution. The quality of the reconstructed frame is measured by the multi-scale SSIM (MS-SSIM) index [36], which has good performance in coherence with human perception.

The rest of the paper is organized as follows. The background of optimal bit allocation and related work are reviewed in Section II. Then, the problems in block-level bit allocation are discussed in Section III. In Section IV, the maximum distortion descend (MDD) algorithm is designed for block-level bit allocation, and a simplified and practical framework for video coding is proposed. The experimental results are given in Section V, and the discussions are presented in Section VI. Finally, we conclude the paper in Section VII.

## II. BACKGROUND AND RELATED WORKS

In this section, we introduce the background of bit allocation optimization, and discuss the influence of dependency on bit allocation, then the related works for dependent bit allocation are introduced.

### A. Optimal Bit Allocation

Assume we are using $R_c$ bits to encode $N$ sources (frames or blocks) in a video. $X=\{x_i|i=1,\dots,N\}$ is a candidate scheme of quantization, where $x_i$ is the QP of the $i^{th}$ source. Denote $r_i(X)$ and $d_i(X)$ as the rate and distortion of the $i^{th}$ source, respectively, the total number of bits is $R(X) = \sum_{i=1}^{N} r_i(X)$, and the overall distortion is noted as $D(X)$. The optimal solution of bit allocation can be solved by the following constrained optimization problem:

$$X_{opt} = \arg\min_{X} D(X) \text{ s.t. } R(X) \le R_c. \quad (1)$$

There are two commonly used definitions of $D(X)$, which results in two different optimization problems. One is the MINAVE-based problem, where $D(X) = \sum_{i=1}^{N} d_i(X)/N$. In this case, the optimization minimizes the average distortion of the sources. The other one is the MINMAX-based problem, where $D(X) = \max_{i} d_i(X)$, it focus on minimizing the maximum distortion of the sources.

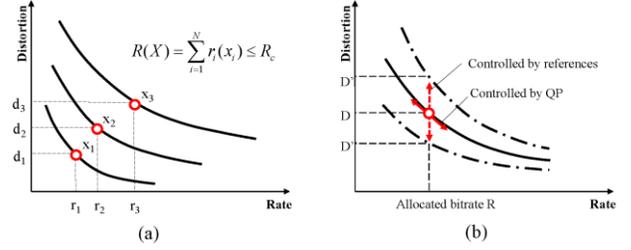
Fig. 1. The RD property of sources and its affection on bit allocation. (a) Independent sources. (b) Dependent sources.

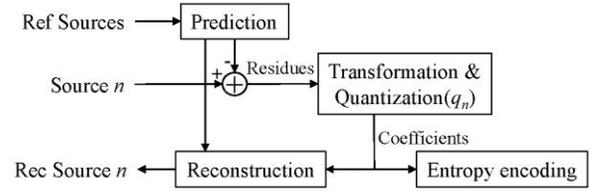
Fig. 2. Framework of the hybrid video encoder.

In independent bit allocation, sources are encoded separately, and their R-D properties are self-determined. Fig. 1(a) shows a simple illustration of bit allocation among three independent sources. The R-D point ($r_i$, $d_i$) moves along the $i^{th}$ source's R-D curve is controlled only by the QP $x_i$. Suppose there are $|x|$ available QPs for each source, then the maximum number of candidate solution of $X$ is $N|x|$. In other words, we need at most $N|x|$ times of encoding to find the optimal solution.

However, in modern hybrid video codecs, intra and inter prediction are used to remove the redundancy between sources. As a result, the sources are not independent with each other anymore. Fig. 2 shows the framework of hybrid video encoder. Source $n$ is first predicted by its reference sources, and only the residues (prediction error) are encoded by QP $x_n$. Generally, a better reference produces less prediction error, thus the residues are affected by the encoding parameters (QPs) of the references. As a result, the final R-D point of source $n$ is not only controlled by its own QP $x_n$, but also affected by the QPs of the references, as it is illustrated in Fig. 1(b). The dependency ties the QP selection of the dependent sources together, which brings great difficulty for optimization. In the worst case, all the sources are tied together (e.g. the MBs in intra frame of H.264), we needs to try all the possible QP combinations to find the optimal solution. That is, totally $|x|^N$ times of encoding are needed. This is unacceptable in practical encoding.

### B. Related Works

Different methods have been developed to solve the optimal bit allocation problem. Lagrange multiplier method is widely used to solve the MINAVE-based independent bit allocation problem [37]. For the dependent problem, DP method is first adopted [16], and then model-based accelerating methods are introduced in practical encoding. Besides, many iteration-based methods are also developed for the MINMAX-based problem in order to achieve constant quality among sources.

*1) Lagrangian Multiplier Method*

Lagrangian multiplier method is usually used to convert the constrained optimization problem in (1) into an unconstrained optimization problem as:

$$X_{opt} = \arg\min_X J(X) \qquad (2)$$

$$= \arg\min_X \left( D(X) + \lambda \cdot \sum_{i=1}^{N} r_i(X) \right), \qquad (3)$$

where, $J(X)$ is the Lagrangian cost function. $\lambda$ is the Lagrangian multiplier determined by $R_c$, which can be derived by the bisection method [38]. For the independent MINAVE problem, this problem can be optimized separately on each source as follows:

$$\min_X J(X) = \sum_{i=1}^{N} \min_{x_i} \left( d_i(x_i) + \lambda \cdot r_i(x_i) \right). \qquad (4)$$

The Lagrangian method needs to know all the sources' R-D property in advance. However, the R-D property of a dependent source could not be known before its reference sources are encoded (refer to Fig. 1(b)). As a result, the Lagrangian method could not be used in dependent bit allocation directly.

*2) Dynamic Programming (DP) Method*

The DP method is usually used to solve the MINAVE-based dependent bit allocation problem [16-19], in which the optimization is to find a shortest path, with QPs as it nodes and the Lagrangian cost as its length. It has denoted by [17] that the computation complexity of DP under a given $\lambda$ is $O(N|x|^M)$, where $M$ is the maximum number sources dependent with each other, and $M=1$ corresponds to the independent case. When $M=N$, the DP method reverts to the full search method which has complexity of $O(|x|^N)$. The high complexity of DP prevents it from being used in the block-level bit allocation.

*3) Model-based Methods*

Some other works developed accelerating methods for the MINAVE-based dependent problem. Since the obstacle for using Lagrangian multiplier method is the unachievable R-D property of the dependent sources before actually encoding, these methods predict the R-D property of a dependent source based on the encoding results of its reference source [20-27]. Though the optimal solution is not guaranteed, this is due to the inaccuracy of the models, these methods can greatly decrease the computational complexity and yield promising results. However, all these methods are confined to the frame-level bit allocation, and based on the assumption that the R-D property of two adjacent frames are similar. In block-level bit allocation, the R-D property of two adjacent blocks may differ greatly due to their different contents. Besides, the R-D property of a dependent block usually depends on several blocks but not a single one. As a result, the existing model-based accelerating methods could not be used in block-level bit allocation.

*4) Methods for MINMAX-based problem*

MINMAX promises that no single source will be excessively distorted, as a result, the quality of sources will be nearly consistent at the given constraint. The DP method can be used for the MINMAX-based bit allocation [17], but also suffers for the high computational complexity. Some other works [29-32] use multi-pass methods to pursue constant quality among frames. Among these methods, the multistage method [31] can achieve nearly constant quality among frames. It contains two stages: the target rate stage and the target distortion stage. The target rate stage makes sure the total used bits meet the rate constraint, and the target distortion stage forces all the sources to be encoded to the same distortion. Iteration is needed to find the solution that meets the constraint. These methods have much less encoding complexity than the DP method, and usually be used in offline applications.

However, all these methods do not care about the quality fluctuation between blocks within a frame. This is mainly because in these methods the quality is measured by mean square error (MSE). Since the contents in contiguous frames are very similar, pursuing the constant MSE among frames can lead to constant perceptual quality. It is quite different in the block level: the contents of blocks can vary greatly within a frame, and enforcing the same MSE among blocks is meaningless. Besides, the strict constant quality constraint is difficult to be used in block-level bit allocation, which will be discussed in detail in the next section.

III. DIFFICULTIES IN BLOCK-LEVEL BIT ALLOCATION

Though there are many practical methods to optimize the frame-level bit allocation, it is difficult to use them in block-level. The difficulties of block-level bit allocation are discussed in this section. Due to these difficulties, it is necessary to develop new methods that suitable for block-level bit allocation.

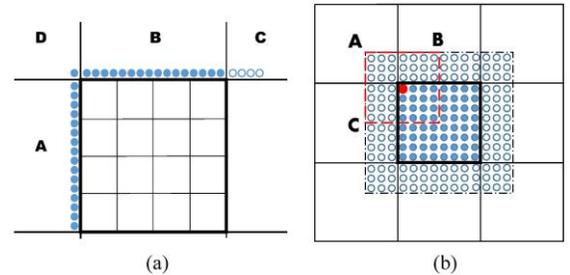

Fig. 3. MB-level dependency. (a) Dependency between intra MBs. (b) Dependency of perceptual distortion measurement between MBs.

*A. Dependency between Intra Blocks*

The intra prediction used in hybrid video coders brings strong dependency between the blocks in intra frame. Fig. 3(a) illustrates the structure of intra prediction in H.264. Where the bold solid box is the current encoding MB. In intra16x16 mode prediction, the pixels in the adjacent encoded MBs 'A', 'B', and 'D' noted by the solid dots are used in prediction. The pixels noted by the hollow dots in MB 'C' are further used in intra4x4 mode prediction. As a result, the encoding of the current MB depends on the encoding of MBs 'A'~'D'. Since each MB in the intra frame depends on the encoding of MBs before it (except the first one), all the intra MBs are tied together. As a result, the DP method is too complex to be used. Besides, since each MB depends on multi reference MBs, the model-based method could not be used either.

## B. Dependency Introduced by Distortion Measurement

The most commonly used metric to measure the distortion of an MB is the mean square error (MSE), which is defined as:

$$d_{MSE} = \frac{1}{K}\sum_{i=1}^{K}(x_i - y_i)^2, \quad (5)$$

where $x_i$ and $y_i$ are the $i^{th}$ pixel in the original and reconstructed block, respectively, and $K$ is the number of pixels in an MB. $d_{MSE}$ is calculated independently, no pixels of the other blocks are needed. However, when using the perceptual-based IQA metrics, sliding windows are usually used in calculation, and this brings dependency between blocks. Fig. 3(b) illustrates the calculation of the SSIM index. To calculate the SSIM index of the MB noted by the bold solid box, a sliding window (the red dash box) centered at the $i^{th}$ pixel is used to calculate a SSIM value $SSIM(x_i, y_i)$ between the original and reconstructed MB. Then the average value at each pixel in the MB is used as the SSIM index of the MB been discussed. The SSIM-based distortion can be defined as:

$$d_{SSIM} = 1 - \frac{1}{K}\sum_{i=1}^{K} SSIM(x_i, y_i). \quad (6)$$

It is clear that, the $d_{SSIM}$ of an MB depends on the pixels in the surrounding MBs, which is affected by the QPs of the surrounding MBs. In this sense, the R-D property of a block is not independent, even no intra prediction is used (e.g. the inter MBs in P frames in H.264). It is easy to understand that, all the MBs in a frame are tied together when using $d_{SSIM}$ as distortion metric. As a result, the DP method and model-based method are also inapplicable.

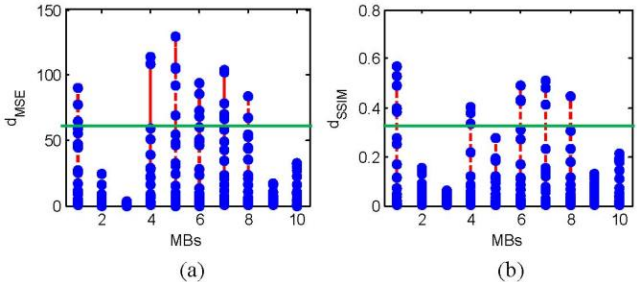

Fig. 4. The reachable distortion of 10 randomly selected MBs in the first intra frame of "*paris*". (a) $d_{MSE}$. (b) $d_{SSIM}$.

## C. Range of Reachable Distortion

In prediction-based encoding, a better reference usually brings better results. As a result, if the prediction of an MB is very good, then the residues would be too small to be distorted. In this sense, the reachable distortion of an MB is limited and may be very different from the others. Fig. 4 shows distortion (measured by $d_{MSE}$ and $d_{SSIM}$, respectively) of 10 randomly selected MBs in the first intra frame of "*paris*" that encoded by H.264 with all possible QPs (from 0 to 51). It is clear that, the maximum reachable distortion of different MBs are quite different. In this case, the multistage method for MINMAX (refer to section II-B) could not be used here. The target distortion stage of the multistage method forces every MB to be encoded into the same distortion, but the target distortion (e.g. the green solid line in Fig. 4) may be unreachable for some MBs. The MBs in inter frames also have this problem.

## IV. MODEL AND ALGORITHM FOR OPTIMAL BLOCK-LAYER BIT ALLOCATION

In this section, we develop an iterative-based method for MB-level bit allocation in H.264. Description of problem and theorems about the MINMAX-based bit allocation are first given, and then a maximum distortion descend (MDD) method is proposed, finally the MDD method is simplified and used in a framework of MB-level bit allocation.

### A. Optimal Bit Allocation based on MINMAX

Due to the discussion in the last section, the constant quality (distortion) solution may not exist in MB-level bit allocation. We can find a solution with minimum variation of distortion among sources instead of the strict constant quality solution. The variation of distortion among all sources is defined as:

$$V_d(X) = \max_i d_i(X) - \min_j d_j(X). \quad (7)$$

Then, the problem can be formulated as:

$$\min_X V_d(X) \text{ s.t. } \sum_{i=1}^{N} r_i(X) = R_c. \quad (8)$$

We can see that, if there exist $X^o$ such that $V_d(X^o)=0$, then $X^o$ is the constant quality solution. We have the following theorem.

***Theorem 1***: $V_d(X)$ will be minimized under the constraint of $R_c$, if there exist $X^*$ such that $R^* = \sum_{i=1}^{N} r_i(X^*) = R_c$ and:

$$d_i(X^*) = \begin{cases} d_{i,\max}, & \text{if } d_{i,\max} < d^* \\ d^*, & \text{otherwise} \end{cases}, \quad (9)$$

where $d_{i,\max}$ is the maximum reachable distortion of the $i^{th}$ source, and $d^*$ is a scalar.

***Proof***: Please refer to Appendix A.

For a given target distortion $d^\#$, we can find the $X^\#$ such that:

$$d_i(X^\#) = \begin{cases} d_{i,\max}, & \text{if } d_{i,\max} < d^\# \\ d^\#, & \text{otherwise} \end{cases}, \quad (10)$$

and the corresponding used bits $R^\# = \sum_{i=1}^{N} r_i(X^\#)$. We have the following theorem, which leads to an iterative algorithm to find the optimal solution.

***Theorem 2***: $R^\#$ is a non-increasing function of $d^\#$.

***Proof***: Please refer to Appendix B.

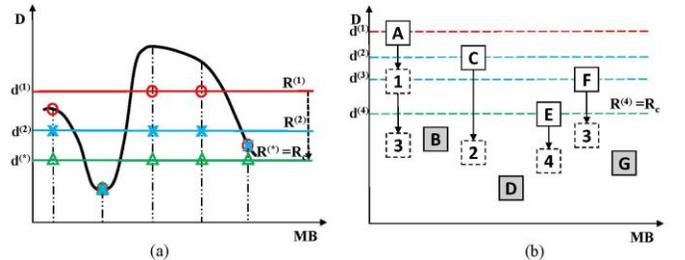

Fig. 5. Illustration of the MDD method. (a) The iterative process to find the optimal solution. (b) An example of the MDD process.

### B. Maximum Distortion Descend (MDD) Algorithm

Based on the above two theorems, we can iteratively find the optimal solution under the bit constraint $R_c$. Since $R^\#$ is a non-increasing function of $d^\#$ (*Theorem* 2), and the $X^\#$ corresponds to $d^\#$ is the optimal solution with the constraint $R^\#$ (*Theorem* 1), we can carefully adjust $d^\#$ until $R^\#=R_c$, then $X^\#$ is the optimal solution corresponds to $R_c$. Fig. 5(a) illustrates this process (where the black curve states for the maximum achievable distortion of the sources): try a large enough target distortion $d^{(1)}$ first, the corresponding quantization scheme $X^{(1)}$ results in $d_i^{(1)}$ (the dots noted by the red 'O') and $R^{(1)}$; if $R^{(1)}<R_c$, we try a smaller target distortion $d^{(2)}$ until we find the $d^{(*)}$ that $R^{(*)}=R_c$, then the corresponded $X^{(*)}$ (which results in the $d_i^{(*)}$ noted by the green '△') is the optimal solution.

To realize the above idea, we propose an iterative maximum distortion descend (MDD) algorithm, as it is shown in Algorithm 1. Where, $QP_{\max}$ is the maximum permitted QP for a source. The process of MDD can be illustrated by an example in Fig. 5(b). The solid boxes represent the distortion of the sources after the initial encoding, and the dashed boxes show the distortion of the sources after their QPs are changed. The number $k$ on the dashed box denotes that the QP is changed in the $k^{\text{th}}$ iteration. In each iteration, the target distortion is set as the current maximum distortion. Based on *Theorem* 2, the resulted number of bits $R^{(k)}$ increases with the decrease of the target distortion $d^{(k)}$, and approaches $R_c$ gradually. We can find that, some sources' QP may never be changed throughout the iteration, e.g. sources 'B', 'D', and 'G'.

---

**Algorithm 1.** Maximum distortion descend (MDD) algorithm
**Input:** $R_c$
**Output:** $X$
**begin**
  *Initial encoding:* $X=\{x_i=QP_{\max}|i=1,\ldots,N\}$, $r_i(X)$, $d_i(X)$.
  $j=\arg\max_{i=1,\cdots,N} d_i(X)$.
  $k=1$, $d^{(k)}=\max_{i=1,\cdots,N} d_i(X)$, $R^{(k)}=\sum_{i=1}^{N} r_i(X)$.
  **while** $R^{(k)}<R_c$
    $x_j=x_j-1$.
    *Encoding:* update $X$, $r_i(X)$, $d_i(X)$.
    $j=\arg\max_{i=1,\cdots,N} d_i(X)$.
    $k=k+1$, $d^{(k)}=\max_{i=1,\cdots,N} d_i(X)$, $R^{(k)}=\sum_{i=1}^{N} r_i(X)$.
  **end**
**end**

---

### C. Complexity of MDD

Now we can estimate the complexity of the MDD algorithm. In the independent case, suppose there are $|x|$ permitted QPs for each source, since we only need to encode one source in each iteration and the maximum number of QP changing for a source is $|x|$, the maximum number of needed source encoding is $N|x|$, thus the complexity in independent encoding is O($N|x|$). In the dependent case, the maximum number of needed source encoding in each iteration is $N$, thus the complexity of dependent encoding is O($N^2|x|$), which is much lower than the DP-based methods.

However, the complexity of the MDD method is still too high for practical encoding. In practice, it is not necessary to find the exact optimal solution, this allows us to do some simplification on the MDD method.

### D. Simplification of MDD

Some simplifications can be taken on MDD method to accelerate the optimization. 1) We do not need to use $QP_{\max}$ in the initial encoding, a large enough QP is acceptable. 2) More than one source's QP can be changed in each iteration. 3) The change step of QP can be bigger than 1. The simplified method of MDD is shown in Algorithm 2. Where, the initial QP $QP_0$ is selected based on the bit constraint $R_c$: by encoding all the sources with $QP_0$, the total number of bits should be close enough to $R_c$. In model-based rate control methods, $QP_0$ can be easily estimated by an R-Q model with enough accuracy. $\theta$ is a positive integer, which is used to control the initial distortion of the sources. $\alpha$ is a float number between 0 to 1, which is used to prevent $R$ to be larger than $R_c$. The function sort(D) sorts the sources in order of descending distortion, and returns the index of the sources. $L$ controls how many sources' QP can be changed in each iteration, and $\delta$ is the changing step of QP.

---

**Algorithm 2.** Simplified MDD
**Input:** $R_c$, $QP_0$
**Output:** $X$
**begin**
  *Initial encoding:* $X=\{x_i=QP_0+\theta|i=1,\ldots,N\}$, $r_i(X)$, $d_i(X)$.
  $R=\sum_{i=1}^{N} r_i(X)$.
  **while** $R<R_c\times\alpha$
    $D=\{d_i(X)|i=1,\ldots,N\}$.
    $S=\text{sort}(D)$,
    $T=\{S(t)|t=1,\ldots,L\}$.
    $\forall j \in T$, $x_j=x_j-\delta$.
    *Encoding:* update $X$, $r_i(X)$, $d_i(X)$.
    $R=\sum_{i=1}^{N} r_i(X)$.
  **end**
**end**

---

Though it is defined for the rate constrained ($R_c$) bit allocation, the MDD method can be easily changed to adapt the quality constrained problem. If we want to encode a frame with a quality not lower than $Q_c$, we can just replace the terminating condition $R<R_c\times\alpha$ in Algorithm 2 by $Q>Q_c+\beta$. Where $Q$ is the frame quality of the last iteration (in this paper, it is the MS-SSIM index of the frame), and $\beta$ is a float number between 0 to 1, which is used to prevent $Q$ to be lower than $Q_c$.

### E. MB-level bit allocation (MB-MDD)

Based on the MDD algorithm, we propose an MB-level bit allocation method, named MB-MDD, which determines the QP for each MB in a video sequence under given constraint. Assume we are using $R_T$ bits to encode a GOP, which contains $G$ frames. The framework is shown in Algorithm 3. Where, ***R*** is the set of target bit allocation for each frame, which can be determined by practical frame-level allocation methods, such as the adaptive rate control method recommended by JVT-G012 [39]. ***QP*** is the set of initial QP for each frame corresponds to ***R***, which can be estimated by R-Q models. It should be noted that the distortion of MBs are calculated by $d_{SSIM}$.

**Algorithm 3.** MB-level bit allocation (MB-MDD)
**Input:**   $R_T$
**Output:**  $X$
**begin**
  *frame-level allocation:*
    $R = \{R_{c,j} | j=1,\ldots,G\}$, where $\sum_{j=1}^{G} R_{c,j} = R_T$.
    $QP = \{QP_{0,j} | j=1,\ldots,G\}$
  *MB-level allocation:*
    **for** $i=1,\ldots,G$
      $X_j = \text{MDD}(R_{c,j}, QP_{0,j})$.
    **end**
    $X = \{X_j | j=1,\ldots,F\}$.
**end**

## V. EXPERIMENTAL RESULTS

The proposed MB-MDD method is implemented based on H.264/AVC reference software JM18.6 [40]. The parameters in the MDD method are empirically set as: $\theta$ =3, $\alpha$ =-5e-5, $\beta$ =0.97, and $\delta$ =2. The coding configuration are set as follows: all intra and inter modes are enabled; one reference frames; each GOP has 15 frames, with one I frame and follows by 14 P frames; high complexity RDO.[1]

### A. Experiments Setup

Since there is no existing method of optimal MB-level QP selection for H.264/AVC, here we evaluate the proposed method against the fixed-QP method when rate control is off in JM18.6, and the rate control method recommended by JVT-G012. For JVT-G012, the size of BU is set to a single MB. To be fair, we compare the proposed method with the two competing methods under either the same bitrate or the same quality. For this, we use the results of the competing methods as the constraint of the MB-MDD method.

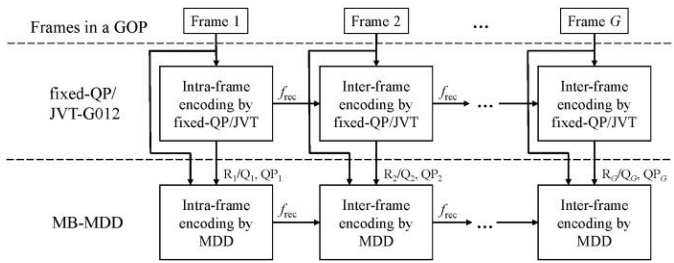

Fig. 6. Encoding process of the experiments.

The encoding process is shown in Fig. 6, where $G$ is the number of frames in a GOP, $R_i$ and $Q_i$ are the number of bits and the quality of the $i^{\text{th}}$ frame encoded by the fixed-QP or JVT-G012 method, respectively, $QP_i$ is the initial QP, $f_{\text{rec}}$ is the reconstructed frame which is used as reference for the encoding of the next frame. The encoding is called "rate constrained" when using $R_i$ as the constraint for MDD, and "quality constrained" when using $Q_i$.

The encoding performance are compared at both high bitrate and low bitrate. For the fixed-QP method, two QP sets are used: $QP_H$={18, 22, 26} and $QP_L$={26, 30, 34}. For the JVT-G012 method, we use different target bitrates for different sequences (see Table 1 for details), the high and low bitrate corresponds to $R_H$={$R_1$, $R_2$, $R_3$} and $R_L$={$R_3$, $R_4$, $R_5$}, respectively.

Table 1. Five levels ($R_1$~$R_5$) of target bitrate (Mbps) used for JVT-G012 on different sequences

| Sequence | $R_1$ | $R_2$ | $R_3$ | $R_4$ | $R_5$ |
|---|---|---|---|---|---|
| *container*(CIF), *news*(CIF), *silent*(CIF) | 1.2 | 1 | 0.8 | 0.6 | 0.4 |
| *foreman*(CIF), *paris*(CIF) | 1.8 | 1.5 | 1.2 | 0.9 | 0.6 |
| *bus*(CIF), *coastguard*(CIF), *flower*(CIF), *mobile*(CIF), *stefan*(CIF), *tempete*(CIF), *walk*(CIF) | 3.6 | 3 | 2.4 | 1.8 | 1.2 |
| *mobcalter*(720P) | 45 | 35 | 25 | 15 | 5 |
| *parkjoy*(720P), *duckstakeoff*(720P) | 75 | 60 | 45 | 30 | 15 |

### B. Performance Evaluation on Intra frame Encoding

We first evaluate the encoding performance of MB-MDD on intra frames. Since the fixed-QP and JVT-G012 methods encode intra frames in the same way, we just use the results of the fixed-QP method for comparing.

#### 1) Quality Constrained Encoding

With quality constrained, the bits spent by the two methods are compared in Table 2. d$Q$ and d$R$ state for the quality improvement and the percentage of bitrate saving by MB-MDD over the fixed-QP method, respectively. The MS-SSIM index is used to measure the perceptual quality of reconstructed frames. $N_{\text{iter}}$ is the average number of iteration that needed for each frame by MB-MDD. d$V$ states for the reduction of average quality fluctuation. The average quality fluctuation of an encoded sequence is calculated by averaging the standard deviation of the MB distortions of individual frames, and the distortion of MB is measured by $d_{\text{SSIM}}$.

Table 2 shows that, MB-MDD can save a lot of bitrate for intra frame encoding (10% in average, up to 15.5%), while on most sequences, MB-MDD achieves a little higher quality than fixed-QP. We can also see that, the MB-MDD method can greatly reduce the quality fluctuation inside frames (more than 40% in average, up to 60.3%). The proposed method needs less than 10 iterations in average for each frame.

#### 2) Rate Constrained Encoding

Fig. 7 shows the rate-quality curves of the first intra frame of CIF sequences "*bus*" and "*paris*". Clearly, with the same bitrate to fixed-QP, MB-MDD gets distinctly better frame quality. In Figs. 8 and 9, we zoom-in the encoded outputs by the two methods (when QP=34 for fixed-QP). We can see that MB-MDD preserves much better fine details of texture areas.

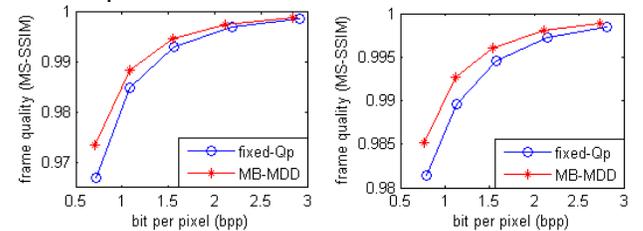

Fig. 7. Rate-quality curves of the intra frames in "*bus*" (left) and "*paris*" (right) encoded by MB-MDD and fixed-QP.

---
[1] Examples of encoded videos by the proposed method and supplementary materials can be found at http://gr.xjtu.edu.cn/web/xqmou/downloads/cpq.

Table 2. Comparison between MB-MDD and fixed-QP on intra frames in quality constrained encoding.

| Sequences | $QP_H=\{18, 22, 26\}$ | | | | $QP_L=\{26, 30, 34\}$ | | | |
|---|---|---|---|---|---|---|---|---|
| | $dQ$(1e-4) | $dR$ | $dV$ | $N_{iter}$ | $dQ$(1e-4) | $dR$ | $dV$ | $N_{iter}$ |
| *bus*(CIF) | 0.138 | 11.7% | 39.8% | 7.8 | 7.431 | 13.4% | 38.8% | 8.2 |
| *coastguard*(CIF) | 0.773 | 5.6% | 50.9% | 10.3 | 5.895 | 6.9% | 43.0% | 10.6 |
| *container*(CIF) | 4.684 | 11.1% | 21.0% | 8.5 | 7.885 | 15.5% | 21.2% | 7.0 |
| *flower*(CIF) | -0.276 | 10.8% | 38.4% | 5.8 | 1.332 | 12.2% | 41.5% | 6.8 |
| *foreman*(CIF) | 1.847 | 9.5% | 38.9% | 8.8 | 4.879 | 9.7% | 37.1% | 8.3 |
| *mobile*(CIF) | -0.167 | 7.8% | 57.6% | 9.8 | 1.051 | 6.7% | 60.3% | 11.0 |
| *news*(CIF) | 0.854 | 12.0% | 34.0% | 8.0 | 2.634 | 11.0% | 42.8% | 7.8 |
| *paris*(CIF) | 0.423 | 12.2% | 38.9% | 7.6 | 3.799 | 13.7% | 48.5% | 7.8 |
| *silent*(CIF) | 2.507 | 7.8% | 48.4% | 9.0 | 7.355 | 6.4% | 33.3% | 9.4 |
| *stefan*(CIF) | 0.116 | 13.6% | 37.4% | 8.3 | 1.297 | 10.8% | 53.6% | 11.6 |
| *tempete*(CIF) | 0.527 | 9.8% | 55.6% | 9.1 | 1.763 | 9.6% | 50.8% | 9.9 |
| *walk*(CIF) | 0.617 | 12.3% | 31.5% | 8.4 | 3.455 | 10.1% | 40.9% | 8.3 |
| *parkjoy*(720P) | 3.184 | 11.6% | 42.2% | 8.3 | 18.575 | 14.8% | 42.1% | 8.9 |
| *mobcalter*(720P) | 1.892 | 11.7% | 46.2% | 8.2 | 4.146 | 11.4% | 41.9% | 9.4 |
| *duckstakeoff*(720P) | 0.333 | 7.6% | 57.9% | 9.8 | 3.267 | 6.0% | 52.2% | 10.4 |
| **Average** | 1.163 | 10.3% | 42.6% | 8.5 | 4.984 | 10.6% | 43.2% | 9.0 |

Table 3. Comparison between MB-MDD and fixed-QP on the whole sequence in quality constrained encoding.

| Sequences | $QP_H=\{18, 22, 26\}$ | | | | $QP_L=\{26, 30, 34\}$ | | | |
|---|---|---|---|---|---|---|---|---|
| | $dQ$(1e-4) | $dR$ | $dV$ | $N_{iter}$ | $dQ$(1e-4) | $dR$ | $dV$ | $N_{iter}$ |
| *bus*(CIF) | -0.131 | 19.1% | 49.6% | 9.6 | 2.072 | 29.0% | 54.1% | 11.4 |
| *coastguard*(CIF) | 1.164 | 7.0% | 50.7% | 10.5 | 6.401 | 10.6% | 41.6% | 10.9 |
| *container*(CIF) | 2.691 | 14.8% | 19.4% | 7.7 | 4.961 | 12.5% | 20.6% | 6.3 |
| *flower*(CIF) | -0.046 | 16.8% | 39.3% | 5.7 | 0.917 | 20.9% | 38.7% | 8.0 |
| *foreman*(CIF) | 0.793 | 12.7% | 44.7% | 10.3 | 2.353 | 10.9% | 39.9% | 9.6 |
| *mobile*(CIF) | -0.172 | 11.6% | 60.6% | 10.5 | 0.420 | 12.2% | 65.9% | 12.6 |
| *news*(CIF) | 0.218 | 11.4% | 32.1% | 6.2 | 0.830 | 1.3% | 41.2% | 7.6 |
| *paris*(CIF) | 0.090 | 15.8% | 39.3% | 7.5 | 1.366 | 18.1% | 47.8% | 9.3 |
| *silent*(CIF) | 0.948 | 3.1% | 45.7% | 7.2 | 2.913 | -2.6% | 32.4% | 7.7 |
| *stefan*(CIF) | 0.176 | 19.4% | 46.9% | 10.6 | 0.738 | 10.7% | 56.9% | 12.6 |
| *tempete*(CIF) | 0.191 | 13.1% | 61.1% | 10.3 | 1.166 | 14.5% | 53.5% | 11.2 |
| *walk*(CIF) | 0.896 | 15.7% | 34.8% | 8.6 | 3.228 | 16.2% | 45.0% | 9.1 |
| *parkjoy*(720P) | 0.937 | 22.6% | 59.6% | 11.7 | 3.755 | 36.6% | 62.0% | 11.9 |
| *mobcalter*(720P) | 1.286 | 18.0% | 45.6% | 8.2 | 2.235 | 3.1% | 39.7% | 8.9 |
| *duckstakeoff*(720P) | 0.546 | 6.6% | 57.3% | 10.0 | 4.934 | 4.2% | 53.3% | 10.7 |
| **Average** | 0.639 | 13.8% | 45.8% | 9.0 | 2.552 | 13.2% | 46.2% | 9.9 |

Table 4. Comparison between MB-MDD and JVT-G012 on the whole sequence in quality constrained encoding. The target bit rates ($R_1$~$R_5$) are defined in Table 1 for different sequences.

| Sequences | $R_H=\{R_1, R_2, R_3\}$ | | | | $R_L=\{R_3, R_4, R_5\}$ | | | |
|---|---|---|---|---|---|---|---|---|
| | $dQ$(1e-4) | $dR$ | $dV$ | $N_{iter}$ | $dQ$(1e-4) | $dR$ | $dV$ | $N_{iter}$ |
| *bus*(CIF) | 0.371 | 17.8% | 41.0% | 8.6 | 0.599 | 18.5% | 50.2% | 11.4 |
| *coastguard*(CIF) | 1.057 | 10.7% | 54.3% | 10.5 | 3.069 | 11.0% | 47.5% | 10.6 |
| *container*(CIF) | 2.109 | 14.0% | 19.8% | 8.2 | 2.266 | 15.0% | 21.4% | 8.1 |
| *flower*(CIF) | -0.123 | 9.1% | 31.4% | 5.3 | 0.125 | 7.9% | 34.3% | 6.7 |
| *foreman*(CIF) | 0.466 | 13.4% | 41.4% | 9.2 | 0.678 | 13.6% | 44.3% | 10.9 |
| *mobile*(CIF) | -0.038 | 11.7% | 59.5% | 11.0 | 0.206 | 10.1% | 62.5% | 11.8 |
| *news*(CIF) | 0.071 | 11.9% | 36.0% | 7.3 | 0.263 | 9.9% | 46.2% | 9.0 |
| *paris*(CIF) | 0.002 | 15.5% | 37.3% | 8.4 | 0.347 | 16.6% | 40.7% | 9.0 |
| *silent*(CIF) | 0.467 | 7.6% | 54.2% | 9.0 | 1.240 | 8.0% | 47.1% | 9.2 |
| *stefan*(CIF) | 0.014 | 10.7% | 35.0% | 9.2 | 0.154 | 6.9% | 48.4% | 12.7 |
| *tempete*(CIF) | 0.052 | 13.7% | 58.4% | 9.9 | 0.308 | 13.4% | 62.0% | 11.6 |
| *walk*(CIF) | 0.467 | 18.5% | 31.3% | 8.1 | 1.314 | 18.1% | 37.3% | 9.0 |
| *parkjoy*(720P) | 0.273 | 13.3% | 42.8% | 8.4 | 0.909 | 17.1% | 51.0% | 10.7 |
| *mobcalter*(720P) | 1.000 | 25.2% | 50.4% | 8.3 | 1.734 | 22.8% | 49.2% | 9.4 |
| *duckstakeoff*(720P) | 0.178 | 7.3% | 60.5% | 10.4 | 1.748 | 6.9% | 59.1% | 10.9 |
| **Average** | 0.425 | 13.4% | 43.6% | 8.8 | 0.997 | 13.1% | 46.8% | 10.1 |

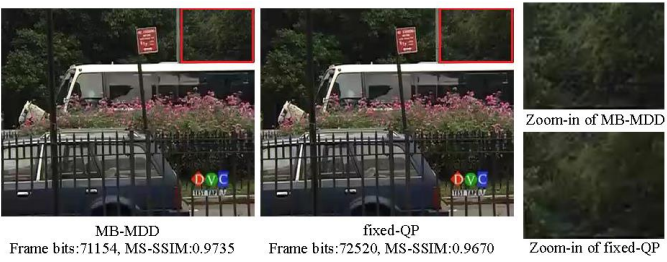

Fig. 8. Visual comparison of the encoded intra frame of "*bus*" between MB-MDD and fixed-QP (QP=34).

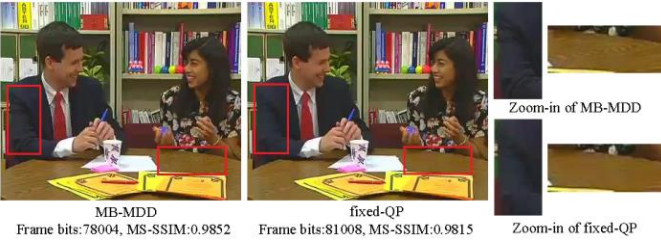

Fig. 9. Visual comparison of the encoded intra frame of "*paris*" between MB-MDD and fixed-QP (QP=34).

*3) SSIM-MDD vs. MSE-MDD*

In the above experiments, SSIM was used as the quality metric, and we call it SSIM-MDD. Here we perform experiments to evaluate MDD with non-IQA metrics by replacing $d_{SSIM}$ with $d_{MSE}$, which is called MSE-MDD. Fig. 10 shows the rate-quality curves of the first intra frame of sequences "*bus*" and "*paris*" in rate constrained encoding. Clearly, MSE-MDD has even lower frame quality than fixed-QP. The reason will be discussed in Section V-A. Fig. 11 shows the reconstructed frames of "*paris*" encoded by SSIM-MDD and MSE-MDD. The constraint of bitrate is set by fixed-QP with QP=30. One can easily find that SSIM-MDD preserves much more frame details than MSE-MDD while using nearly the same number of bits.

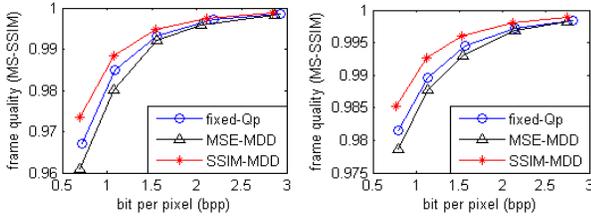

Fig. 10. Rate-quality curves of the intra frames in "*bus*" (left) and "*paris*" (right) encoded by fixed-QP, SSIM-MDD and MSE-MDD.

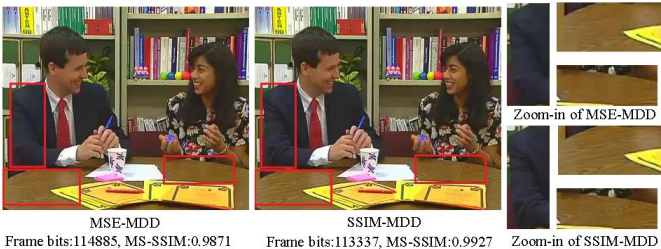

Fig.11. Visual comparison of the encoded intra frame of "*paris*" between MSE-MDD and SSIM-MDD in rate constrained encoding.

### C. Performance Evaluation on the Whole Sequence

In this sub-section, we evaluate the encoding performance on the whole sequence.

*1) Quality Constrained Encoding*

Tables 3 and 4 present the average quality improvement and bitrate saving by MB-MDD over fixed-QP and JVT-G012, respectively. From Table 3, we can see that the bitrate of almost all the sequences are greatly reduced (13% in average, up to 36.6%) by the proposed method, while the quality of them are kept almost unchanged or even slightly improved. Table 4 shows that MB-MDD achieves better performance than JVT-G012 on almost all the sequences under the given bit rate (13% bitrate saving in average, up to 25.2%). In both Table 3 and Table 4, the average quality fluctuation is greatly reduced. We can also find that, the effectiveness of MB-MDD on different sequences are different, this will be discussed in the next section.

*2) Rate Constrained Encoding*

The rate-quality curves on some sequences by the MB-MDD compared with the fixed-QP and JVT-G012 methods in rate constrained encoding are shown in Figs. 12 and 13, respectively. One can clearly see that MB-MDD can improve the image perceptual quality obviously under the same bitrate. Fig. 14 use the first three GOP of "*bus*" as example to compare the quality fluctuation on each frame between the MDD method and the fixed-QP method when QP=30. We can find that the MDD method can greatly reduce the quality fluctuation on each frame in rate constrained encoding.

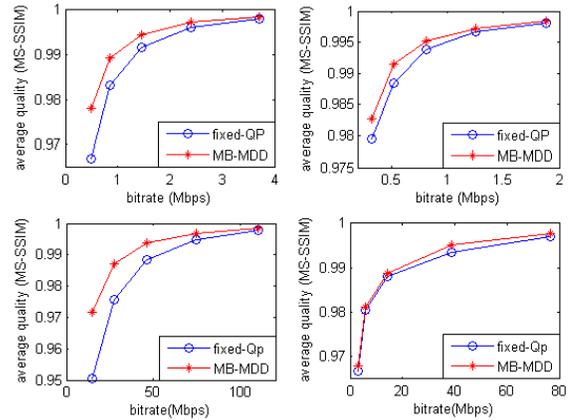

Fig. 12. Rate-quality curves of sequences (from left to right and top to bottom) "*bus*", "*paris*", "*parkjoy*" and "*mobcalter*" encoded by MB-MDD and fixed-QP.

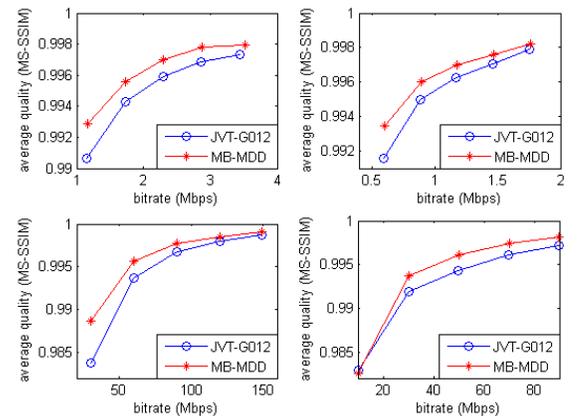

Fig. 13. Rate-quality curves of sequences (from left to right and top to bottom) "*bus*", "*paris*", "*parkjoy*" and "*mobcalter*" encoded MB-MDD and JVT-G012.

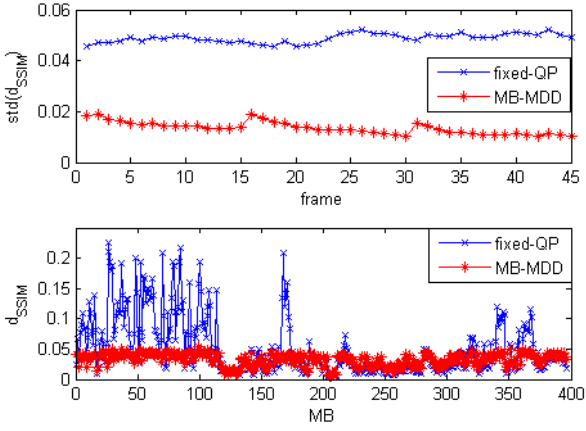

Fig. 14. Comparison of the quality fluctuation of the encoding results between MB-MDD and fixed-QP when QP=30. Up: Standard deviation of $d_{SSIM}$ of each frame in the first 3 GOP of "*bus*". Bottom: the $d_{SSIM}$ of MBs in the 30[th] frame of "*bus*".

### D. Subjective Performance Evaluation

To further verify the perceptual quality improvement brought by the proposed MB-MDD method, we conduct subjective quality evaluation tests based on the two-alternative-forced-choice (2AFC) method, as done by Wang et al. [11] and Yeo et al. [12]. In each trial, a subject is shown a pair of video sequences and is asked (forced) to choose the one he/she thinks to have better quality. We select 6 sequences and encode each sequence with three different settings: the fixed-QP method, quality constrained MDD method (MDD-QC), and rate constrained MDD method (MDD-RC). The rate and quality of the sequences under each setting are shown in Table 5.

We carry out two 2AFC tests. The first test compares the fixed-QP method with MDD-QC, and the second test compares the fixed-QP method with MDD-RC. There are 6 pairs of video sequences in each test. Each sequence pair is played four times with random order, and 10 subjects participate in the experiments. Hence, in each 2AFC test we obtained 240 evaluation outputs: 24 for each subject and 40 for each sequence pair.

Table 5. Quality (MS-SSIM) and bit rate for sequences used in subjective evaluation.

| Sequence | fixed-QP | | MDD-QC | | | MDD-RC | |
|---|---|---|---|---|---|---|---|
| | Quality | Rate (kbps) | Quality | Rate (kbps) | Rate Saved (%) | Quality | Rate (kbps) |
| *bus*(CIF) | 0.965 | 474 | 0.965 | 320 | 32.49 | 0.978 | 474 |
| *paris*(CIF) | 0.989 | 480 | 0.989 | 387 | 19.38 | 0.992 | 480 |
| *container*(CIF) | 0.965 | 194 | 0.966 | 175 | 9.79 | 0.972 | 194 |
| *coastguard*(CIF) | 0.971 | 794 | 0.972 | 658 | 17.13 | 0.975 | 794 |
| *mobcalter*(720P) | 0.968 | 2851 | 0.968 | 2541 | 10.87 | 0.970 | 2851 |
| *parkjoy*(720P) | 0.951 | 15459 | 0.951 | 9630 | 37.71 | 0.972 | 15456 |

The results of the two 2AFC subjective tests are shown in Figs. 15 and 16, respectively. In each figure, we show the proportion of choices that prefer the proposed MDD method over the fixed-QP method. It can be seen from Fig. 15 that the overall proportion preferring MDD is about 60%. It should be noted that MDD costs much less bitrate than fixed-QP. From Fig. 16, one can see that all the subjects are in favor of the sequences encoded by the MDD method. The results validate that compare with the fixed-QP method, the proposed MDD method can achieve better quality at the same bit rate or save bits while maintaining similar (or even better) level of quality.

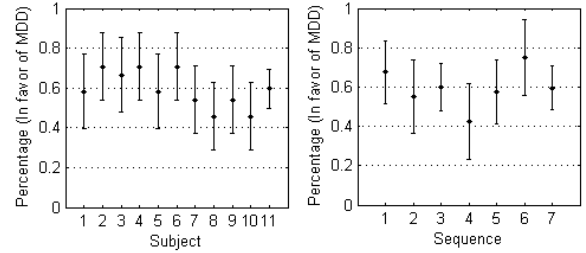

Fig. 15. Error bar plot of subjective test to compare fixed-QP with MDD-QC. Left: preference for individual subject (1~10: subject number, 11: average). Right: preference for individual sequence (1~6: sequence number, 7: average).

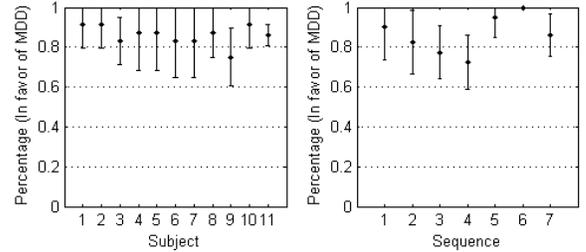

Fig. 16. Error bar plot of subjective test to compare fixed-QP with MDD-RC. Left: preference for individual subject (1~10: subject number, 11: average). Right: preference for individual sequence (1~6: sequence number, 7: average).

## VI. DISCUSSIONS

In the previous sections, we developed a framework to optimize bit allocation at MB-level to make the perceptual quality of MBs as consistent as possible. The experimental results showed that compared with fixed-QP and JVT-G012, the MDD method can save more than 10% bit rates in average while ensuring the quality (MS-SSIM) of frames. Both the subjective and objective evaluations demonstrated the advantage of MDD. In this section, we discuss more about the performance of MDD from two aspects, the quality metric and the content of video sequence.

### A. Quality Metric for MDD

It has been shown in Section IV-A that MDD can achieve better encoding performance than fixed-QP when $d_{SSIM}$ is used as the distortion measure. However, when $d_{MSE}$ is used to measure the distortion of blocks, MDD does not work well. To explain why MSE is not a good quality metric for MDD, we analyze the MDD encoding process by using the first frame of sequence "*paris*". The illustration is shown in Fig. 17.

The initial state of MDD is the output of fixed-QP encoding. Then the distortion measurement of MBs determines to what direction the QPs should be adjusted. Fig. 17(b) shows the encoded frame by fixed-QP, and Figs. 17(e) and (f) show its distortion maps measured by $d_{MSE}$ and $d_{SSIM}$, respectively. It can be found that the SSIM metric is more adaptive to the image content than the MSE metric. The distortions of image fine scale structures (e.g., MB *B*) caused by quantization are detected by SSIM but ignored by MSE. In contrast, MSE penalizes more in the areas with high brightness, which often have small perceptual distortions (e.g., MB *A*). Let's compare the distortions of two typical MBs in Fig. 17(c) and (d). Clearly, the encoded MB *A* has much better perceptual quality than the encoded MB *B*. However, the distortion index MSE gives the opposite answer: MB *A* has larger $d_{MSE}$ (22.58) than MB *B* (4.13), which wrongly indicates that MB*B* has better quality than MB *A*. On the contrary, the distortion index $d_{SSIM}$ is in accordance to human observation, and MB*A* has lower $d_{SSIM}$ (0.0189) than MB *B* (0.0907). In order to reduce quality fluctuation between MBs, the QPs for MB*B* should be decreased to reduce the distortion. Unfortunately, by using $d_{MSE}$ to measure the quality, the QPs for MB *A* will be decreased, leading to wrong QP adjustment.

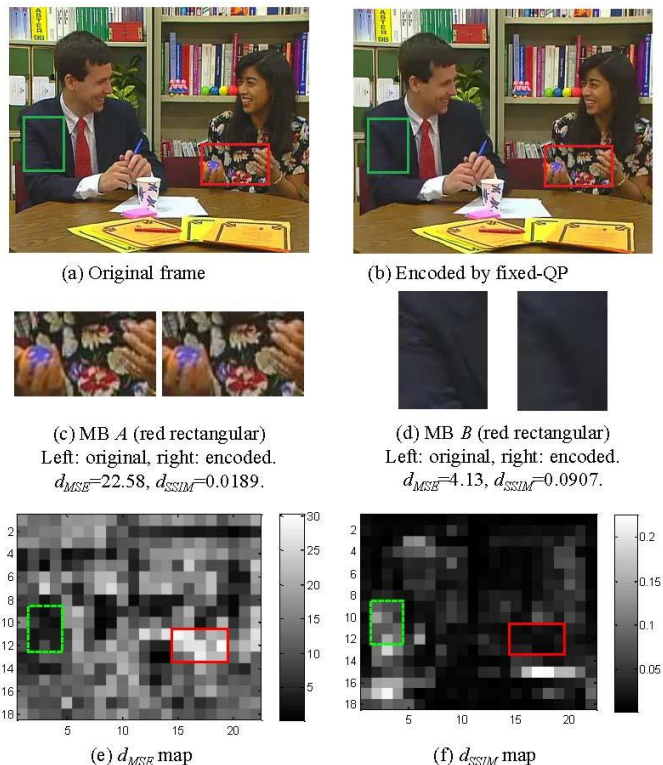

(a) Original frame  (b) Encoded by fixed-QP

(c) MB *A* (red rectangular)
Left: original, right: encoded.
$d_{MSE}$=22.58, $d_{SSIM}$=0.0189.

(d) MB *B* (red rectangular)
Left: original, right: encoded.
$d_{MSE}$=4.13, $d_{SSIM}$=0.0907.

(e) $d_{MSE}$ map   (f) $d_{SSIM}$ map

Fig. 17. The effect of quality metrics on MB-MDD encoding.

### B. The content of sequence

From the experimental results in the last section, we can see that MDD improves the encoding quality on most sequences, but it has different performances on different sequences. The MDD method has very good performance on sequences like "*bus*", "*paris*", and "*parkjoy*", but has little effect on sequences like "*silent*" and "*duckstakeoff*", and on some sequences, like "*coastguard*" and "*mobile*", it only performs well on P frames. It is not difficult to find that the effectiveness of the proposed method is related with the contents of the sequences. Sequences like "*bus*" have both complex scene and a lot of movement, but sequences like "*silent*" have simple scene and nearly no motion. Other sequences like "*coastguard*" have moderately complex scene but with a great degree of movement.

This phenomenon can be explained as follows. MDD aims to reduce the distortions of MBs from an initial state under the constraint of a given bitrate or quality. However, the QPs of some MBs may not be changed since their initial distortions are already very low and they cannot release redundant bits for other MBs to reduce the quality fluctuation in a frame. If there are many such MBs in a frame, it would be hard to save bits by the proposed MDD method. In general, an MB will have a low initial distortion if it can be well predicted by the neighboring MBs. Those frames with simpler scenes and/or smaller degree of motions are more likely to have such MBs, and hence the sequence with such frames will not have big encoding improvement by MDD.

## VII. CONCLUSION

We proposed an MB-level bit allocation method to achieve nearly constant perceptual quality among MBs in a frame. With the MINMAX criterion, an iterative encoding algorithm, namely maximum distortion descend (MDD), was proposed to find solution of the bit allocation constrained by either the number of bits or the frame quality. The proposed MB-MDD framework can effectively solve the MB-level bit allocation problem for perceptual video coding, which is difficult to solve by existing methods, while any existing frame-layer bit allocation method can be used in MB-MDD for its frame-layer control. Both the objective and subjective experimental results showed that the proposed method can greatly improve the encoding performance: either save bits while maintaining the frame quality, or improve the frame quality while using the same number of bits.

### APPENDIX A. PROOF OF THEOREM 1

Let $\Omega=\{i|d_i(X^*)=d^*\}$, and $\Psi=\{i|d_i(X^*)=d_{i,max}\}$. Assume that there exists an *X'* such that $R'=\sum_{i=1}^{N} r_i(X')=R_c$, and the corresponding distortions of MBs are $d_i(X')$, $i=1,\ldots,N$. From the definition of $d_{i,max}$, we have $d_i(X') \leq d_i(X^*)=d_{i,max}$, $\forall i \in \Psi$.

1) If $d_i(X')=d_i(X^*)=d_{i,max}$, $\forall i \in \Psi$, then $\sum_{i \in \Psi} r_i(X') = \sum_{i \in \Psi} r_i(X^*)$. Since $R'=R_c=R^*$, we have $\sum_{i \in \Omega} r_i(X') = \sum_{i \in \Omega} r_i(X^*)$.
   a) If $r_i(X')=r_i(X^*)$, $\forall i \in \Omega$, then $d_i(X')=d_i(X^*)$, $\forall i \in \Omega$, Thus $d_i(X')=d_i(X^*)$, $\forall i$, and $V_d(X')=V_d(X^*)$;
   b) Else, it is certain that $\exists k \in \Omega$, $r_k(X')<r_k(X^*)$. According to the monotonicity of RD performance of each MB (we call this property m.r.d. hereafter), $d_k(X') \geq d_k(X^*)=d^*$, and thus $V_d(X') \geq \max_{i \in \Omega} d_i(X') - \min_j d_{j,max} \geq d^* - \min_j d_{j,max} = V_d(X^*)$.

2) Else if $\exists k \in \Psi$, $d_i(X')<d_i(X^*)$, then $\sum_{i \in \Psi} r_i(X') \geq \sum_{i \in \Psi} r_i(X^*)$. Since $R'=R_c=R^*$, then $\sum_{i \in \Omega} r_i(X') \leq \sum_{i \in \Omega} r_i(X^*)$, and thus $\exists k \in \Omega$, $r_k(X')<r_k(X^*)$. According to the m.r.d., $d_k(X') \geq d_k(X^*)=d^*$, we have $V_d(X') \geq \max_{i \in \Omega} d_i(X') - \min_j d_{j,max} \geq d^* - \min_j d_{j,max} = V_d(X^*)$.

In summary, any *X'* that meets the condition $\sum_{i=1}^{N} r_i(X')=R_c$ will have a quality fluctuation of $V_d(X') \geq V_d(X^*)$. Hence, *X\** is the optimal solution to minimizing the quality fluctuation under the constraint of $R_c$.

### APPENDIX B. PROOF OF THEOREM 2

Assume that $d'<d^\#$, and the corresponding vector of QP is *X'*. The MBs can be classified into three sets, and we denote them as $\Omega=\{i|d_i(X')=d'$ and $d_i(X^\#)=d^\#\}$, $\Psi=\{i|d_i(X')=d_{i,max}$ and $d_i(X^\#)=d_{i,max}\}$, and $\Phi=\{i|d_i(X')=d'$ and $d_i(X^\#)=d_{i,max}\}$. According to the m.r.d., since $d'<d^\#$, then $\forall i \in \Omega$, $r_i(X') \geq r_i(X^\#)$, and we have $\forall i \in \Psi$, $r_i(X')=r_i(X^\#)$ and $\forall i \in \Phi$, $r_i(X') \geq$

$r_i(X^{\#})$. Therefore, we have $R'=\sum_{i=1}^{N} r_i(X') \geq \sum_{i=1}^{N} r_i(X^{\#})=R^{\#}$. This means that $R^{\#}$ decreases monotonically with the increase of $d^{\#}$.

REFERENCES

[1] T. Berger and J. D. Gibson, "Lossy source coding," Information Theory, IEEE Transactions on, vol. 44, pp. 2693-2723, 1998.
[2] T. Wiegand, G. J. Sullivan, G. Bjontegaard, and A. Luthra, "Overview of the H.264/AVC video coding standard," *IEEE Transactions on Circuits and Systems for Video Technology*, vol. 13, no. 7, pp. 560–576, 2003.
[3] G. J. Sullivan, J. R. Ohm, W. J. Han, and T. Wiegand, "Overview of the High Efficiency Video Coding (HEVC) Standard," Ieee Transactions on Circuits and Systems for Video Technology, vol. 22, pp. 1649-1668, Dec 2012.
[4] Z. He and S. Mitra, "Optimum bit allocation and accurate rate control for video coding via rho-domain source modeling," *IEEE Trans. Circuits Syst. Video Technol.*, vol. 12, no. 10, pp. 840–849, Oct. 2002.
[5] N. Kamaci, Y. Altinbasak, and R. M. Mersereau, "Frame bit allocation for H.264/AVC video coder via Cauchy density-based rate and distortion models," *IEEE Trans. Circuits. Syst. Video Technol.*, vol. 15, no. 8, pp. 994–1006, Aug. 2005.
[6] S. Ma, W. Gao, and Y. Lu, "Rate-distortion analysis for H.264/AVC video coding and its application to rate control," *IEEE Trans. Circuits. Syst. Video Technol.*, vol. 15, no. 12, pp. 1533–1544, Dec. 2005.
[7] D. K. Kwon, M. Y. Shen, and C. C. J. Kuo, "Rate control for h.264 video with enhanced rate and distortion models," *IEEE Transactions on Circuits and Systems for Video Technology*, vol. 17, no. 5, pp. 517–529, 2007.
[8] A. Ortega and K. Ramchandran, "Rate-distortion methods for image and video compression," *IEEE Signal Processing Magazine*, vol. 15, no. 6, pp. 23–50, 1998.
[9] G. J. Sullivan and T. Wiegand, "Rate-distortion optimization for video compression," Signal Processing Magazine, IEEE, vol. 15, pp. 74-90, 1998.
[10] Y. H. Huang, T. S. Ou, P. Y. Su, and H. H. Chen, "Perceptual rate distortion optimization using structural similarity index as quality metric," *IEEE Transactions on Circuits and Systems for Video Technology*, vol. 20, no. 11, pp. 1614–1624, 2010.
[11] S. Q. Wang, A. Rehman, Z. Wang, S. W. Ma, and W. Gao, "SSIM motivated rate-distortion optimization for video coding," *IEEE Transactions on Circuits and Systems for Video Technology*, vol. 22, no. 4, pp.516–529, 2012.
[12] C. Yeo, H. L. Tan, and Y. H. Tan, "On rate distortion optimization using SSIM," *Circuits and Systems for Video Technology, IEEE Transactions on*, vol. 23, no. 7, pp. 1170–1181, 2013.
[13] S. Q. Wang, A. Rehman, Z. Wang, S. W. Ma, and W. Gao, "Perceptual video coding based on SSIM-inspired divisive normalization," *IEEE Transactions on Image Processing*, vol. 22, no. 4, pp. 1418–1429, 2013.
[14] Z. Mai, C. Yang, K. Kuang, and L. Po, "A novel motion estimation method based on structural similarity for H.264 inter prediction", *IEEE International Conference on Acoustics, Speech and Signal Processing*, 2006.
[15] T. S. Ou, Y. H. Huang, and H. H. Chen, "SSIM-based perceptual rate control for video coding," *IEEE Transactions on Circuits and Systems for Video Technology*, vol. 21, no. 5, pp. 682–691, 2011.
[16] K. Ramchandran, A. Ortega, and M. Vetterli, "Bit allocation for dependent quantization with applications to multi resolution and MPEG video coders," *IEEE Trans. Image Process.*, vol. 3, no. 5, pp. 533–545, Sep. 1994.
[17] G. M. Schuster, G. Melnikov, and A. K. Katsaggelos, "A review of the minimum maximum criterion for optimal bit allocation among dependent quantizers," *IEEE Transactions on Multimedia*, vol. 1, no. 1, pp. 3–17, 1999.
[18] J. Wen, M. Luttrell, and J. Villasenor, "Trellis-based r-d optimal quantization in H.263," *IEEE Trans. Image Process.*, vol. 9, no. 8, pp. 1431–1434, Aug. 2000.
[19] Y. Yang and S. S. Hemami, "Generalized rate-distortion optimization for motion-compensated video coders," *IEEE Transactions on Circuits and Systems for Video Technology*, vol. 10, no. 6, pp. 942–955, 2000.
[20] L. J. Lin and A. Ortega, "Bit-rate control using piecewise approximated rate-distortion characteristics," *IEEE Trans. Circuits. Syst. Video Technol.*, vol. 8, no. 4, pp. 446–459, Aug. 1998.
[21] S. H. Jang and N. Jayant, "An efficient bit allocation algorithm in dependent coding framework and one-way video applications," in *Proc. ICME*, 2004, pp. 383–386.
[22] Y. Sermadevi and S. S. Hemami, "Efficient bit allocation for dependent video coding," *ser. IEEE Data Compression Conference*. Los Alamitos: IEEE Computer Soc., pp. 232–241, 2004.
[23] J. Liu, Y. Cho, Z. Guo, and C. C. Jay Kuo, "Bit Allocation for Spatial Scalability Coding of H.264/SVC With Dependent Rate-Distortion Analysis," *IEEE Trans. Circuits. Syst. Video Technol.*, vol. 20, no. 7, pp. 967–981, Jul. 2010.
[24] S. Hu, S. Kwong, T. Zhao, and C. C. Jay Kuo, "Rate control optimization for temporal-layer scalable video coding," *IEEE Trans. Circuits. Syst. Video Technol.*, vol. 21, no. 8, pp. 1152–1162, Aug. 2011.
[25] C. Pang, O. C. Au, F. Zou, J. Dai, X. Zhang, and W. Dai, "An Analytic Framework for Frame-Level Dependent Bit Allocation in Hybrid Video Coding," *IEEE Trans. Circuits. Syst. Video Technol.*, vol. 23, no. 6, pp. 990–1002, Jun. 2013.
[26] C. Pang, O. C. Au, J. Dai, and F. Zou, "Dependent Joint Bit Allocation for H.264/AVC Statistical Multiplexing Using Convex Relaxation," *IEEE Trans. Circuits. Syst. Video Technol.*, vol. 23, no. 8, pp. 1334–1345, Aug. 2013.
[27] S. Wang, S. Ma, S. Wang, D. Zhao, and W. Gao, "Rate-GOP Based Rate Control for High Efficiency Video Coding," IEEE JOURNAL OF SELECTED TOPICS IN SIGNAL PROCESSING, VOL. 7, NO. 6, DECEMBER 2013.
[28] J. M. Henderson, P. A. Weeks Jr, and A. Hollingworth, "The effects of semantic consistency on eye movements during complex scene viewing," Journal of experimental psychology: Human perception and performance, vol. 25, p. 210, 1999.
[29] M. F. Ringenburg, R. E. Ladner, and E. A. Riskin, "Global MINMAX inter frame bit allocation for embedded video coding," *ser. IEEE Data Compression Conference*, pp. 222–231, 2004.
[30] H. B. Yin, X. Z. Fang, L. Chen, and J. Hou, "A practical consistent quality two-pass VBR video coding algorithm for digital storage application," *IEEE Transactions on Consumer Electronics*, vol. 50, no. 4, pp.1142–1150, 2004.
[31] N. Cherniavsky, G. Shavit, M. F. Ringenburg, R. E. Ladner, and E. A. Riskin, "Multistage: A MINMAX bit allocation algorithm for video coders," *IEEE Transactions on Circuits and Systems for Video Technology*, vol. 17, no. 1, pp. 59–67, 2007.
[32] K. Wang and J. W. Woods, "Mpeg motion picture coding with long-term constraint on distortion variation," *IEEE Transactions on Circuits and Systems for Video Technology*, vol. 18, no. 3, pp. 294–304, 2008.
[33] Z. Wang, A. C. Bovik, H. R. Sheikh, and E. P. Simoncelli, "Image quality assessment: From error visibility to structural similarity," *IEEE Transactions on Image Processing*, vol. 13, no. 4, pp. 600–612, 2004.
[34] Chao Wang, Xuanqin Mou, and Lei Zhang, "Block-Layer Optimal Bit Allocation Based On Constant Perceptual Quality," *Proc. IS&T/SPIE Electronic Imaging 2012 Visual Information Processing and Communication III*, California, USA, 2012.
[35] Chao Wang, Xuanqin Mou, Wei Hong and Lei Zhang, "Block-Layer Bit Allocation for Quality Constrained Video Encoding Based on Constant Perceptual Quality," *SPIE EI 2013 Visual Information Processing and Communication IV (Conference 8666)*, California, USA, 2013.
[36] Z. Wang, E. P. Simoncelli, and A. C. Bovik, "Multi-scale structural similarity for image quality assessment," *ser. IEEE Conference Record of the Asilomar Conference on Signals, Systems and Computers.* New York, pp. 1398–1402, 2003.
[37] Y. Shoham and A. Gersho, "Efficient bit allocation for an arbitrary set of quantizers," *IEEE Transactions on Acoustics, Speech and Signal Processing*, vol. 36, no. 9, pp. 1445–1453, 1988.
[38] C. F. Gerald and P. O. Wheatley, *Numerical analysis*. Addison-Wesley, 2003.
[39] Z. G. LI, F. Pan, K. P. Lim, and G. N. Feng, "Adaptive basic unit layer rate control for JVT," in *Joint Video Team of ISO/IEC MPEG and ITU-T VCEG, JVT-G012, 7th Meeting*, Pattaya, Thailand, 2003, pp. 7–14, JVT-G012.
[40] JVT, "H.264/AVC reference software JM18.6," *[online]* http://iphome.hhi.de/suehring/tml/, 2014.